\documentclass[%
 reprint,
 amsmath,amssymb,
 aps,
]{revtex4-2}

\usepackage{graphicx}
\usepackage{dcolumn}
\usepackage{bm}
\usepackage{xcolor}
\usepackage[english]{babel}
\usepackage{hyperref}
\usepackage{siunitx}
\hypersetup{
    colorlinks=true,
    linkcolor=blue,
    citecolor=blue,
    urlcolor=blue,
}
\usepackage[capitalize]{cleveref}

\usepackage{orcidlink}
\DeclareUnicodeCharacter{2082}{2}

\usepackage{color}

\usepackage[svgnames]{xcolor}

\renewcommand{\vec}[1]{\mathbf{#1}}

\begin{document}

\preprint{APS/123-QED}

\title{Microscopic origin of Rashba coupling from first principles: Layer-resolved orbital asymmetry in transition metal dichalcogenides}

\author{Miguel Morales Cócera$^{1,2}$}
\thanks{\orcidlink{0009-0006-3143-468X} \href{https://orcid.org/0009-0006-3143-468X}{0009-0006-3143-468X}} 

\author{Marta Prada$^{1}$}
\thanks{\orcidlink{0000-0003-3845-9012} \href{https://orcid.org/0000-0003-3845-9012}{0000-0003-3845-9012}} 

\author{Franz Fischer$^{1,2}$}
\thanks{\orcidlink{0009-0007-2416-4199} \href{https://orcid.org/0009-0007-2416-4199}{0009-0007-2416-4199}} 

\author{Gabriel Bester$^{1,3}$}
\thanks{\orcidlink{0000-0003-2304-0817} \href{https://orcid.org/0000-0003-2304-0817}{0000-0003-2304-0817}} 
\affiliation{$^{1}$Institute of Physical Chemistry, University of Hamburg, Luruper Chaussee 149, 22761 Hamburg, Germany}
\affiliation{$^{2}$Max Planck Institute for the Structure and Dynamics of Matter, Luruper Chaussee 149, 22761 Hamburg, Germany}
\affiliation{$^{3}$The Hamburg Centre for Ultrafast Imaging, Luruper Chaussee 149, 22761
Hamburg, Germany}

\date{\today}

\begin{abstract}
Spin-orbit coupling in two-dimensional materials gives rise to a Rashba spin splitting when inversion and mirror symmetries are broken, yet its microscopic origin and quantitative characterization in transition metal dichalcogenides remains incomplete. 
Both symmetries are broken in certain bilayer structures, enabling Rashba splittings in the absence of external electric fields.
We determine this zero-field offset and the Rashba parameters that dictate the  spin splitting in the linear regime. Surprisingly, the splitting is substantially smaller in bilayers than in monolayers at typical fields. This is clarified within a perturbative microscopic model,  revealing that the spin splitting results from a competition between internal polarization and interlayer hybridization. We further introduce the orbital polarization imbalance as an order parameter that captures the asymmetry of the valence bands and determines the spin ordering of the Rashba-split states. Our results are both quantitative and qualitative, as they clarify the nature and origin of Rashba coupling in transition metal dichalcogenides. 
\end{abstract}

\maketitle


\section{Introduction}
A novel class of quasi-two-dimensional (2D) materials---such as bilayers of transition metal dichalcogenides (TMDs) or Janus TMD monolayers---exhibits a built-in electric dipole that arises naturally when mirror symmetry ($m_z$) is broken \cite{Picozzi2014,Wang2022_0}. When inversion symmetry ($\mathcal{I}$) is also broken, a zero-field Rashba spin splitting near the $\Gamma$-point arises \cite{rashba59,doi:10.1126/science.abd3230,Molino2023,Weston2022,PhysRevLett.130.146801}. This feature offers a potential pathway toward low-power spintronic applications with non-volatile spin control \cite{Ghiasi2019}. 

TMDs in particular have emerged at the forefront of spin-related applications due to their strong spin-orbit coupling (SOC) \cite{Bihlmayer2022,Szary2023,Bordoloi2024}, and have been extensively studied as hosts of tightly bound charged and neutral excitons \cite{Nielsen2025_Qtrion,Nielsen2025_Janus}, making them ideal platforms for integrating spin, charge, and optical functionalities \cite{Kovalchuk2025,Huang2025,Chen2025,Gish2024,PhysRevB.109.085407,PhysRevLett.134.026901}. Additionally, recent studies indicate that the $\Gamma$-point neighborhood may play a central role in high-energy excitonic processes in few-layer systems \cite{Chen2025-pf}, as this is the key region in the Brillouin zone where interlayer hybridization, valence band maximum, and electric-field effects converge, giving rise to rich excitonic phenomena that were previously overlooked. 

Despite the growing interest in Rashba SOC in TMDs \cite{Cheng2016,PhysRevB.95.165401,Xiang2019,Gupta2021,PhysRevB.109.085425}, a quantitative and qualitative understanding of its intrinsic origin remains only partially understood. Previous studies have characterized Rashba splittings through energy–momentum extrema, providing limited insight into the fundamental coupling mechanisms  \cite{Cheng2016,PhysRevB.95.165401,Gjerding2021,Xiang2019,PhysRevB.100.155408,Gupta2021,PhysRevB.97.155415,PhysRevB.109.085425,Rezavand2021}. 
Here,  we introduce a framework that resolves the Rashba coefficient, $\lambda_\mathrm{R}^n$, which quantifies the system’s response to an applied external field, and the intrinsic orbital field, $E_0^n$, which captures the built-in asymmetry responsible for finite band-dependent spin splitting at zero bias. These quantities provide a physically transparent description of Rashba effects, bridging atomic-scale asymmetry, orbital hybridization and layer polarization.
Our perspective clarifies why certain bilayer states, despite their built-in dipole at zero field, exhibit a reduced spin splitting response compared to monolayers when subjected to an external out-of-plane electric field, and establishes a unified foundation for comparing Rashba coupling across the TMD family.

We perform systematic \textit{ab initio} calculations of monolayer (ML) and homo-bilayer (BL) MX$_2$ (M = Mo, W; X = S, Se, Te) systems, focusing on the top valence bands near the $\Gamma$-point. 
The observed Rashba-originated spin splittings are elucidated in terms of a microscopic model that treats orbital hybridization and atomic SOC perturbatively, and reveals the  relevant atomistic processes that yield Rashba splitting in  TMDs. 
We introduce the orbital polarization imbalance, $\xi^n$, as a central descriptor of Rashba behavior, quantifying the (band-dependent) orbital asymmetry induced by the breaking of the mirror symmetry $m_z$.
This quantity not only governs the spin ordering (sign of the spin gap), but also describes the intrinsic orbital field which impacts the magnitude of Rashba splitting.
Finally, by performing a Wannierization of the DFT Kohn-Sham states, we investigate the role of the in-plane orbitals---specifically the transition metal $d_{xz/yz}$ and chalcogen $p_{x/y}$---as mediators in the first-order Rashba processes essential to the SOC mechanism.
Our analysis reveals that the magnitude of the Rashba splitting arises from a non-trivial competition between the system’s atomic SOC strengths, orbital polarization and band composition.

\section{Methods}

We performed \textit{ab initio} density functional theory (DFT) calculations using the generalized-gradient approximation (GGA) for the exchange correlation functional, as parametrized by Perdew, Burke and Ernzerhof (PBE) \cite{PhysRevLett.77.3865}, and utilize the DFT-D3-BJ implementation \cite{Grimme2010} to treat the van der Waals interaction. To incorporate SOC we employ fully-relativistic, normconserving pseudopotentials \cite{vanSetten2018} with a plane-wave cutoff energy of 100 Ry using the \textsc{Quantum ESPRESSO} package \cite{Giannozzi2009,Giannozzi2017}. An out-of-plane cell dimension of 50 Bohr has been used to suppress the spurious interaction of periodic images. During structural optimization all atomic forces were converged below 10$^{-4}$ Ry/Bohr leading to lattice constants in close agreement with experimental values \cite{PhysRevB.85.033305,Huang2015,Chen2017,Gusakova2017}. The Brillouin zone was sampled using a $\Gamma$-centered $15 \times 15 \times 1$ Monkhorst-grid \cite{PhysRevB.13.5188}.

Additionally, we wannierized the DFT wave functions of the BL system to obtain an orbital resolved Hamiltonian using the \textsc{wannier90} package \cite{Mostofi2008}. We incorporated chalcogen $p$ and transition metal $d$ orbitals, as these represent the leading orbital contributions in the close vicinity of the Fermi level \cite{PhysRevB.88.245436,PhysRevB.86.165108}, leading to a basis set size of 44 Wannier functions. Furthermore a denser $k$-grid ($21 \times 21 \times 1$) allowed reliable results \cite{PhysRevB.56.12847,RevModPhys.84.1419}.


\section{Results}

\subsection{SOC in TMDs: From monolayer to bilayer}
\label{subsec:from-mono}

\begin{figure}[htbp]
    \centering
    \includegraphics[width=0.5\textwidth]{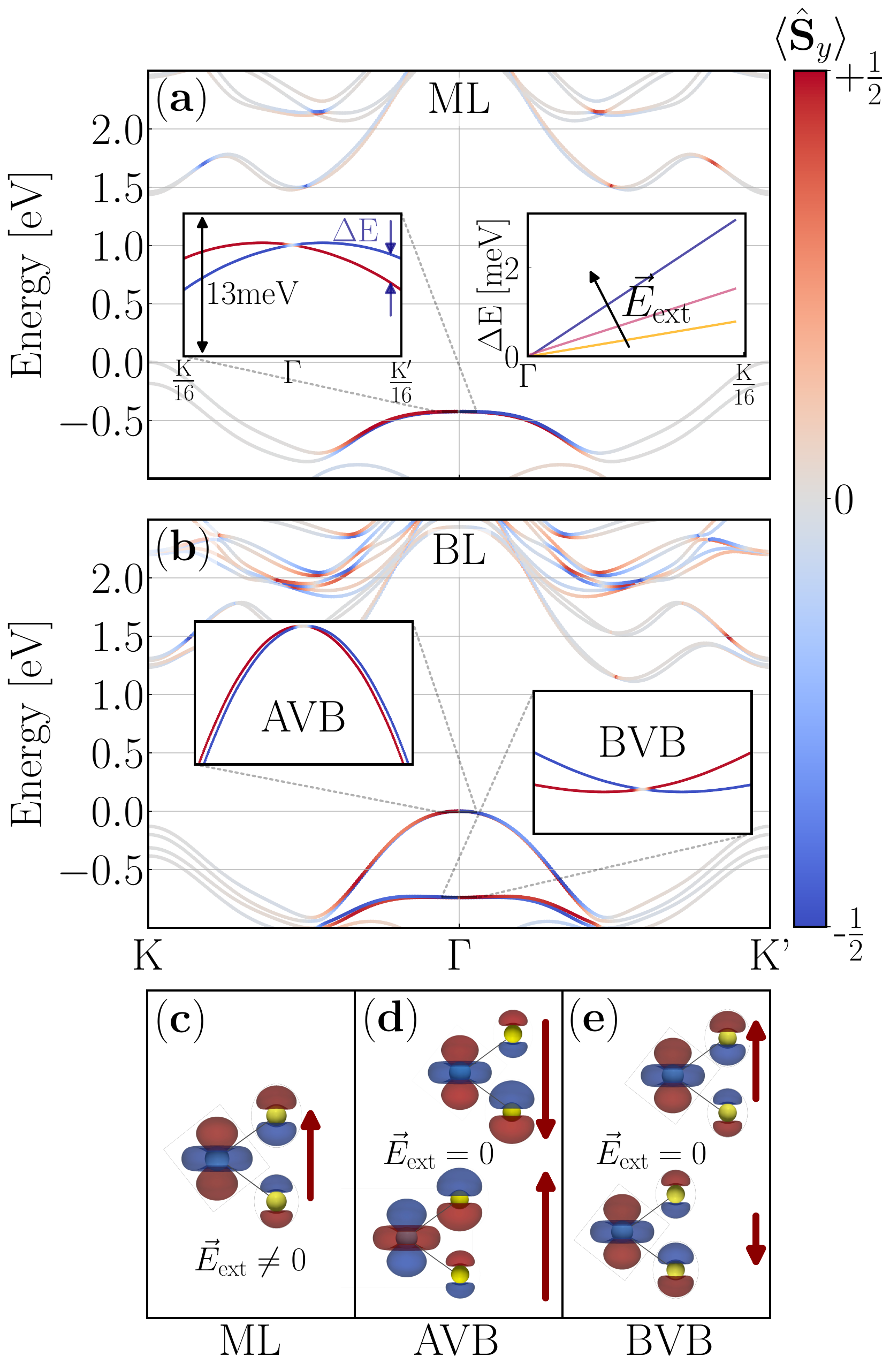}
    \caption{Band structure of MoSe$_2$ with color-coded $\langle \hat{\textbf{S}}_{y} \rangle$ projections for ML (a) and BL (b) in the R$_\mathrm{X}^\mathrm{M}$ stacking. The insets correspond to the zoom-ins of the target bands, with equal energy (13 meV) and momentum window (from ${\mathrm{K}}/{16}$ to ${\mathrm{K}^{\prime}}/{16}$) for quantitative comparison between the states. The right inset in (a) shows the spin splitting $\Delta$E for different electric fields (yellow to blue: 0.05, 0.1 and 0.2 V/$\mathrm{\AA}$). Wave function representation of the top VB of a  ML under an external electric field $E_{\mathrm{ext}}$ (c), and of AVB (d) and BVB (e) in the BL at zero field. The color of the orbitals schematically represents their phase. Red arrows represent the orbital polarization at the ML level.}
    \label{fig:bands_intro.pdf}
\end{figure}

We consider ML and BL  MX$_2$ (M = Mo, W; X = S, Se, Te) TMDs in the 1H phase at the ML level. 
MLs lack inversion symmetry ($\mathcal{I}$) but preserve the mirror symmetry ($m_z$), which forbids any linear-in-$k$ Rashba term; hence, only a small cubic splitting is allowed unless an external electric field breaks $m_z$.
In BLs, $\mathcal{I}$ is restored in the H-stacking, but it is broken in R-type stacking. Among them, only the  R$_\mathrm{X}^\mathrm{M}$  configuration [symmetry equivalent to the  R$_\mathrm{M}^\mathrm{X}$ in homobilayers,  see \cref{fig:bands_intro.pdf}(d),(e)] breaks both  $\mathcal{I}$  and $m_z$ \cite{capelluti13}, enabling Rashba splitting. We therefore focus on MLs under out-of-plane electric fields and on BLs in the  R$_\mathrm{X}^\mathrm{M}$ stacking, also called R$_\mathrm{hm}$, R$_\mathrm{h}^\mathrm{M}$, B$^{\mathrm{M/X}}$ and AB in the literature \cite{Huang2022}.

In \cref{fig:bands_intro.pdf}(a) we show the band structure of MoSe$_2$ along $k_x$ with an out-of-plane electric field $E_{\mathrm{ext}}$ = 0.2 V/$\mathrm{\AA}$. 
The bands are colored according to their spin expectation value $\langle \hat{\textbf{S}}_{y} \rangle$ to highlight the dominant spin component perpendicular to momentum, as expected from the Rashba spin texture~\cite{rashba59,ji22}. 
The left inset shows a magnification of the Rashba spin splitting and the right inset illustrates the linearity of that spin splitting, $\Delta \mathrm{E}$, with respect to $k$ for increasing values of external field (yellow to blue: 0.05, 0.1 and 0.2 V/$\mathrm{\AA}$), which is a feature of the Rashba effect. 
We stress that the small spin splitting at $\Gamma$, we are addressing here, significantly increases along $k_x$ until it reaches the $\mathrm{K}$-point, where the spins are oriented out-of-plane (zero in-plane spin expectation values in \cref{fig:bands_intro.pdf}(a)). This splitting is sometimes referred to as Zeeman-type~\cite{Yuan2013,Wang2022}, and leads to the A and B exciton splitting in optics.

\Cref{fig:bands_intro.pdf}(b) shows the corresponding plot for BL of MoSe$_2$ with R$_\mathrm{X}^\mathrm{M}$ stacking in the absence of external electric field. The insets show the non-zero spin splitting of the topmost two (four, including spin) $\Gamma$-point valence bands.
These topmost VBs are usually explained by a layer hybridization that splits the $\Gamma$-point VB maximum of the ML into two states in the BL: a lower-energy bonding valence band (BVB) and a higher-energy anti-bonding valence band (AVB)~\cite{Li2007}.  


In \cref{fig:bands_intro.pdf}(c--e), we schematically depict the wave functions at $\Gamma$ for the ML VB and the AVB and BVB of the BL. Red arrows indicate the orbital polarization induced by either the external electric field $E_{\mathrm{ext}}$ in case of the ML (c) or the stacking configuration of the BL (d),(e). The phase of the wave functions show the anti-bonding character of AVB and the bonding character of BVB. Interestingly, the bonding state displays less charge accumulation in the interlayer region than the AVB, contrary to typical expectations.

\subsection{Rashba coefficient $\lambda^n_\mathrm{R}$ and intrinsic orbital field $E_0^n$}

ML TMDs belong to the point group $D_{3h}$, and the symmetry reduces to $C_{3v}$ upon application of an out-of-plane electric field, which breaks $m_z$. Similarly the  R$_\mathrm{X}^\mathrm{M}$ stacked BLs inherently lack both $\mathcal{I}$ and $m_z$ inversion plane even in the absence of an external field; consequently, they also belong to the point group $C_{3v}$.
 As a result, both systems exhibit a linear-in-$k$ Rashba spin splitting around the $\Gamma$-point:
\begin{equation}
    H_{\text{R}} = \alpha_\mathrm{R}^n (k_x\sigma_y-k_y\sigma_x),
    \label{eq:Hr1}
\end{equation}
where ${k_{i}}= p_i / \hbar$ is the crystal momentum, $\sigma_j$ are the Pauli matrices and $\alpha_\mathrm{R}^n$ is the state-dependent Rashba prefactor \cite{Winkler2003}, which additionally depends on the strength of the applied electric field. Without loss of generality, we will focus on the $k_x$ momentum direction.

There are two common methods to extract the field- and band-dependent Rashba parameter $\alpha_\mathrm{R}^n$ from DFT calculations. The first is to identify the valence band maximum, which shifts slightly away from $\Gamma$ in the presence of Rashba splitting, hence $\alpha_\mathrm{R}^n$ can be derived from the $k$-point offset and the corresponding energy difference between the maximum and the $\Gamma$-point \cite{Cheng2016,PhysRevB.100.155408}. 
The second approach, used in this work, involves extracting the spin splitting within the linear regime, both in momentum and in electric field. From \cref{eq:Hr1} and along the $\Gamma$--$\mathrm{K}$ direction,  this is:
\begin{equation}
\label{eq:Hr2}
\Delta \mathrm{E} = \alpha_\mathrm{R}^n k_x = \lambda_\mathrm{R}^n (E_{\text{ext}}+E_0^n) k_x.
\end{equation}

This method allows us to extract both the Rashba parameter $\lambda_\mathrm{R}^n$ and the intrinsic orbital field $E_0^n$, which is a (band-dependent) field related to the internal orbital asymmetry that results from the atomic arrangement and the broken $m_z$, whereas $E_\mathrm{ext}$ is the external field. By calculating $\Delta \mathrm{E}$ for several $k_x$-values within the linear regime, the prefactor $\lambda_\mathrm{R}^n(E_\mathrm{ext}+E_0^n)$ can be obtained directly from the linear dependence (see insets of  \cref{fig:bands_intro.pdf}(a)). By varying $E_{\text{ext}}$, the Rashba parameter $\lambda_\mathrm{R}^n$ and $E_0^n$ can be obtained from the linear dependence as well. 

\begin{figure}[htbp]
    \centering
    \includegraphics[width=0.45\textwidth]{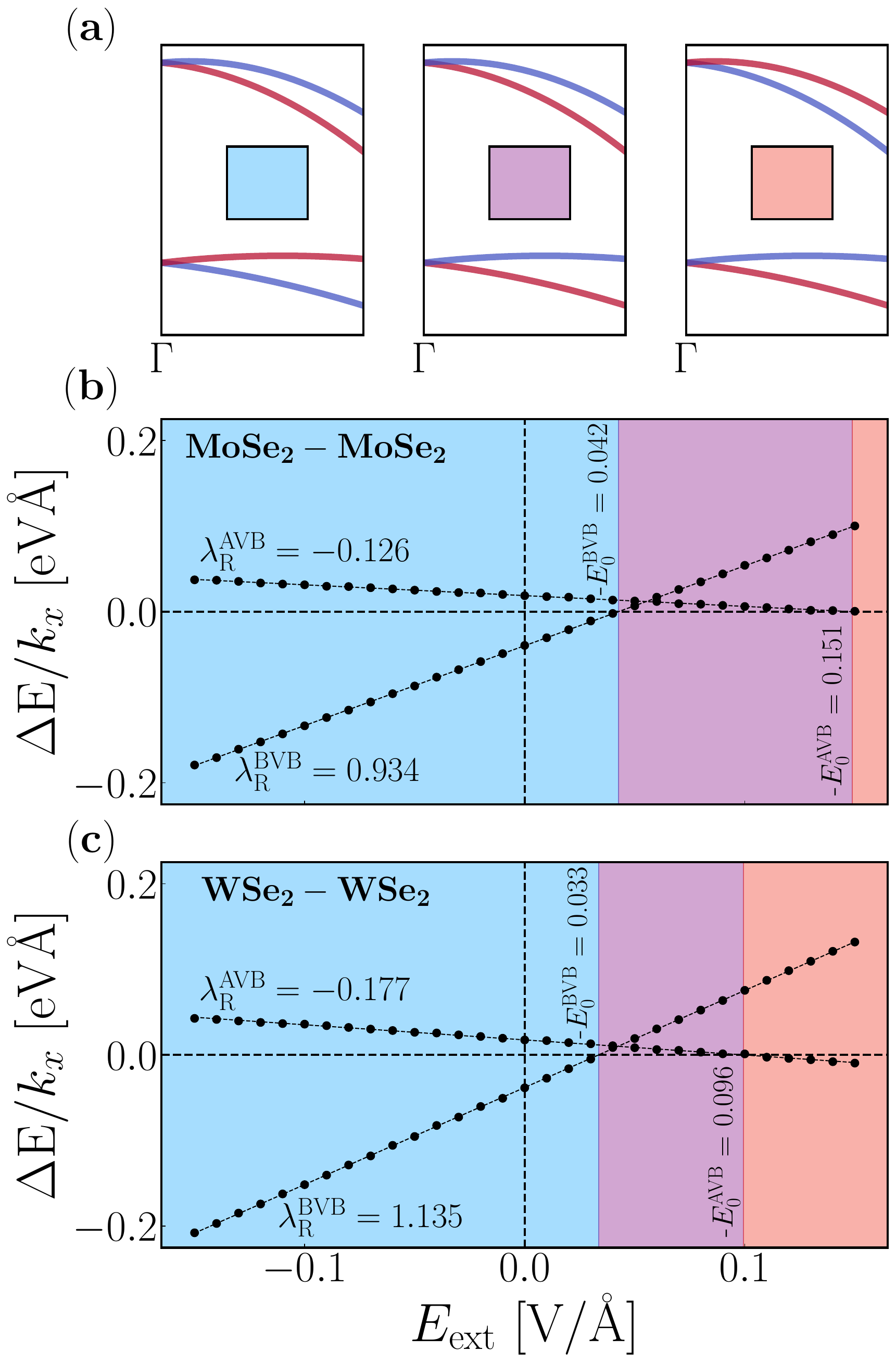}
    \caption{(a) Schematic $\langle \hat{\textbf{S}}_{y} \rangle$-arrangement between spin-split AVB and BVB. The color code defined in (a) is indicated by the background colors in (b) and (c). $\Delta \mathrm{E}/k_x$ as a function of the external field $E_{\mathrm{ext}}$ for AVB and BVB in BL MoSe$_2$ (b) and BL WSe$_2$ (c). Band-dependent Rashba parameters $\lambda_\mathrm{R}^n$ in units of eV$\mathrm{\AA}^2$/V are reported for the relevant states under investigation together with the intrinsic orbital field $E_0^n$ in units of V/$\mathrm{\AA}$.}
    \label{fig:lambda_efield_bil_Se}
\end{figure}

\cref{fig:lambda_efield_bil_Se}(a) illustrates the specific spin textures---defined by the spin expectation value $\langle \hat{\textbf{S}}_{y} \rangle$ along $k_x$---relevant to the spin-split AVB and BVB in BLs. Here, the bands  are colored red for spin-up ($\uparrow$) and blue for spin-down ($\downarrow$). The color codes established in the insets of \cref{fig:lambda_efield_bil_Se}(a) are subsequently used as background indicators in \cref{fig:lambda_efield_bil_Se} for BL MoSe$_2$ (b) and BL WSe$_2$ (c), respectively. Note that the fourth arrangement possibility, AVB with $\uparrow \downarrow$ and BVB with $\uparrow\downarrow$, is not depicted, as it is not observed for the respective BLs. In \cref{fig:lambda_efield_bil_Se}, $\Delta \mathrm{E}/k_x$ is evaluated at $k_x= \mathrm{K}/20$ along the $\Gamma$--$\mathrm{K}$ high-symmetry line as a function of $E_{\mathrm{ext}}$ for both BL systems and their corresponding AVB and BVB. We adopt the convention $\Delta \mathrm{E}= \mathrm{E}_\downarrow - \mathrm{E}_\uparrow$, so that the spin splitting sign is encoded in $\lambda_\mathrm{R}^n$.
Note that $\lambda_\mathrm{R}^n$ is the slope of the splitting with field (linear fit),  while the field at which the splitting vanishes relates to the value of the intrinsic orbital field $E_0^n$:  Namely, the splitting is canceled when the condition $E_{\mathrm{ext}}=-E_0^n$ is met. 

In the absence of an external field, the spin texture arrangement of the AVB is opposite to that of the BVB. The intrinsic field $E_0^n$ signals a reversal of spin character, which is reflected by the change in background color in \cref{fig:lambda_efield_bil_Se}(b) and (c), evidencing the spin control capabilities of the external field.

\begin{figure}[htbp]
    \centering
    \includegraphics[width=0.45\textwidth]{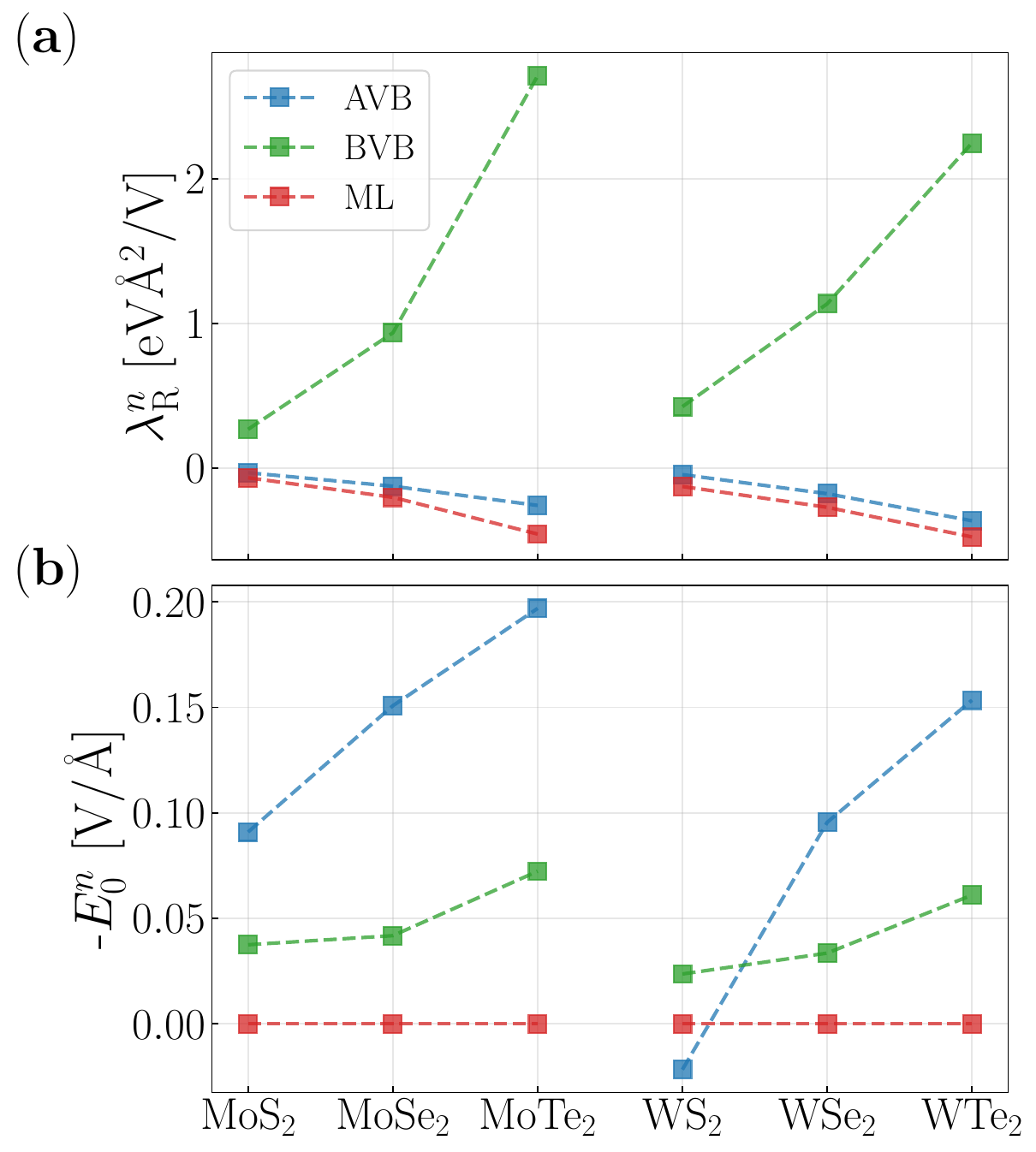}
    \caption{Rashba coefficient $\lambda_\mathrm{R}^n$ (a) and intrinsic orbital field $E_0^n$ (b) of the VB maximum for different TMD MLs, as well as for the AVB and BVB states in the corresponding TMD BLs.}
    \label{fig:rpar_E0}
\end{figure}

In \cref{fig:rpar_E0}(a), we show the Rashba coefficients $\lambda_\mathrm{R}^n$, for different TMDs resolved for the ML VB maximum and the AVB and BVB states of the BL. The values for the AVB are consistently smaller in magnitude than those for the BVB and, notably, also smaller than the corresponding ML values.
Moving to heavier chalcogen atoms (left to right) significantly increases the magnitude of the Rashba coefficients, as the ionicity decreases \cite{Gupta1984}. Recall that less ionic materials have less tendency to bind the charge to the respective atom, hence enhancing their polarizability. As expected, this increase in polarizability goes along with an increase in $\lambda_\mathrm{R}^n$, leading to a stronger response to the applied field. This aligns with the larger magnitude of the intrinsic fields $E_0^n$ as the polarizability (and chalcogen mass) increases, as shown in \cref{fig:rpar_E0}(b). 

On the contrary, the influence of the transition metal on the Rashba parameter appears to be weak. This is related to a competition between the atomistic SOC strength and the orbital polarizability, as the former increases with the atomic mass of the transition metal and the latter decreases. This will be discussed in the following section. A table with the extracted values of $\lambda_\mathrm{R}^n$ and $E_0^n$ for both ML and BL structures of different materials is provided in the Supplementary Material.

\subsection{Orbital polarization imbalance: An order parameter}
\label{subsec:ord-param}

\begin{figure}[htbp]
    \centering
    \includegraphics[width=0.46\textwidth]{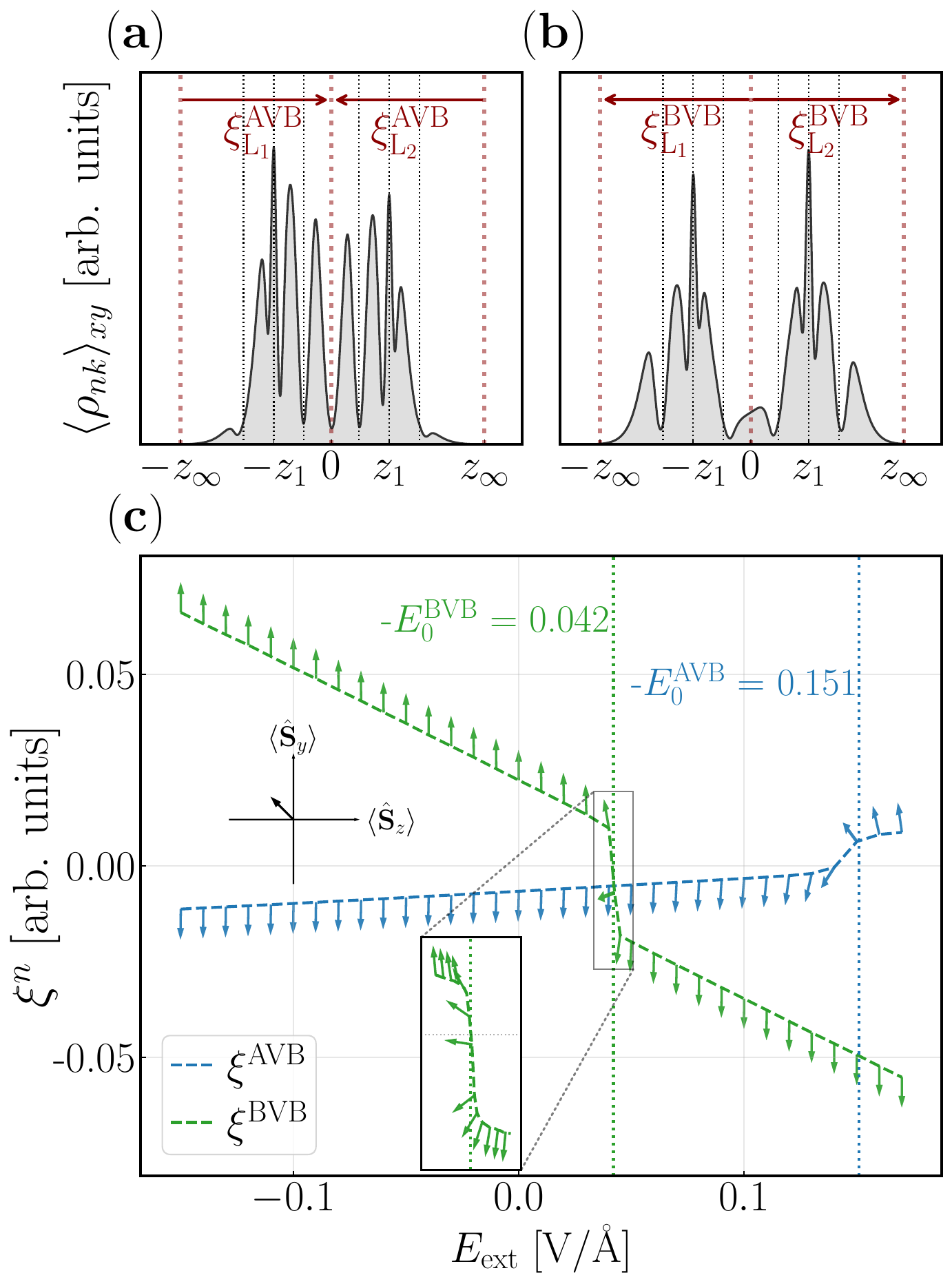}
    \caption{Plane-averaged $\langle \rho_{nk} \rangle_{xy}$ as a function of the out-of-plane coordinate $z$ at $k=\mathrm{K}/40$ for the MoSe$_2$ BL states AVB (a) and BVB (b), respectively. The red arrows indicate the sign (not the magnitude) of the ML $\xi^{n}_{\mathrm{ML}}$ contribution to the BL $\xi^{n}$. Order parameter $\xi^{n}$ for AVB (blue arrows) and BVB (green arrows). The arrows depict the expectation value of the spin of the upper splitted band in the $z$-$y$ plane (see coordinate system). The vertical dotted lines depict the electric field that closes the Rashba splitting in the AVB (blue) and BVB (green) in units of V/$\mathrm{\AA}$.}
    \label{fig:PL_DFT_MoSe2-MoSe2_rho}
\end{figure}

To connect the obtained Rashba spin splitting from \cref{fig:lambda_efield_bil_Se} to a physical observable, we introduce a parameter $\xi^{n}$ that quantifies the uniaxial orbital asymmetry of the $n$-th band at a specific $k$-point. For a ML it takes the form:
\begin{equation*}
\xi^{n}_{\rm ML} =  \left(\int_{-z_\infty}^{z_\mathrm{M}} \langle \rho_{nk} \rangle_{xy}  \, dz- \int_{z_\mathrm{M}}^{z_\infty} \langle \rho_{nk} \rangle_{xy}\, dz\right)C^{-1},  
\end{equation*}
where  $\langle \rho_{nk} \rangle_{xy}$ is the plane-averaged charge density of band $n$ at momentum $k$,  $z_\mathrm{M}$ is the $z$-coordinate of the metal plane, and  $C = \int\langle \rho_{nk} \rangle_{xy} \, dz$ is a normalization constant. In other words, $\xi^{n}_{\rm ML}$ encodes the asymmetry in the charge distribution with respect to the plane of the transition metal. This quantity is zero in the absence of an external electric field, since the ML maintains $m_z$. 
For the BL we define similarly:
\begin{equation*}
\xi^{n}= (\xi_{\mathrm{L}_1}^{n} + \xi_{\mathrm{L}_2}^{n})C^{-1}\quad,
\end{equation*}
with 
\begin{eqnarray}
\xi_{\mathrm{L}_1}^{n} &=&  \int_{-z_\infty}^{-z_1} \langle \rho_{nk} \rangle_{xy} dz \, -\int_{-z_1}^{0} \langle \rho_{nk} \rangle_{xy} dz ,  \nonumber \\
\xi_{\mathrm{L}_2}^{n} &=&  \int_{0}^{z_1} \langle \rho_{nk} \rangle_{xy}  \, dz- \int_{z_1}^{z_\infty} \langle \rho_{nk} \rangle_{xy}\, dz  ,
\label{eq:PL}
\end{eqnarray}
with $\pm z_1$ being the transition metal planes and $z = 0$ the middle point between these planes, as shown in \cref{fig:PL_DFT_MoSe2-MoSe2_rho}(a),(b). 
In the BL, $\xi^{n}$ is somewhat more subtle, as it describes the difference in the MLs asymmetries, or the orbital polarization imbalance. This term can be zero, even if both MLs within the BL have a non-vanishing asymmetry. These ML asymmetries can cancel each other, just as two dipoles can lead to an overall zero dipole.   

In \cref{fig:PL_DFT_MoSe2-MoSe2_rho}(c) we show the orbital polarization imbalance, $\xi^{n}$, for the AVB (blue) and BVB (green) as a function of electric field for MoSe$_2$ for $k = \mathrm{K}/40$. Remarkably, $\xi^n$ exhibits a step at 0.042 V/$\mathrm{\AA}$ and 0.151 V/$\mathrm{\AA}$ for BVB and AVB, respectively. The step occurs precisely at the external fields where the spin splitting $\Delta \mathrm{E}$ is zero, i.e. when $E_{\mathrm{ext}}=-E_0^n$ (see \cref{fig:lambda_efield_bil_Se}(c)). Additionally, the difference in slopes of \cref{fig:lambda_efield_bil_Se}(c) is qualitatively reproduced by the order parameter $\xi^n$. While the absolute values of the slopes of the order parameter depend on the chosen $k$, the qualitative behavior remains consistent (see Supplementary Figures S1 and S2).
The arrows in \cref{fig:PL_DFT_MoSe2-MoSe2_rho}(c) depict the expectation value of spin $\langle \bf \hat {\vec S} \rangle$  of the upper band in the $z$-$y$ plane. The spin projections reveal a clear Rashba spin texture, which disappears at $E_{\mathrm{ext}}=-E_{0}^n$ (see zoom-in of the step in \cref{fig:PL_DFT_MoSe2-MoSe2_rho}(c)). 
The analysis of the orbital polarization imbalance $\xi^n$ demonstrates its robustness as a reliable indicator of a spin splitting across all examined systems (see Supplementary Figure S3). Moreover, it confirms that the Rashba coupling arises from layer-specific orbital asymmetries.

\begin{figure}[htbp]
    \centering
    \includegraphics[width=0.42\textwidth]{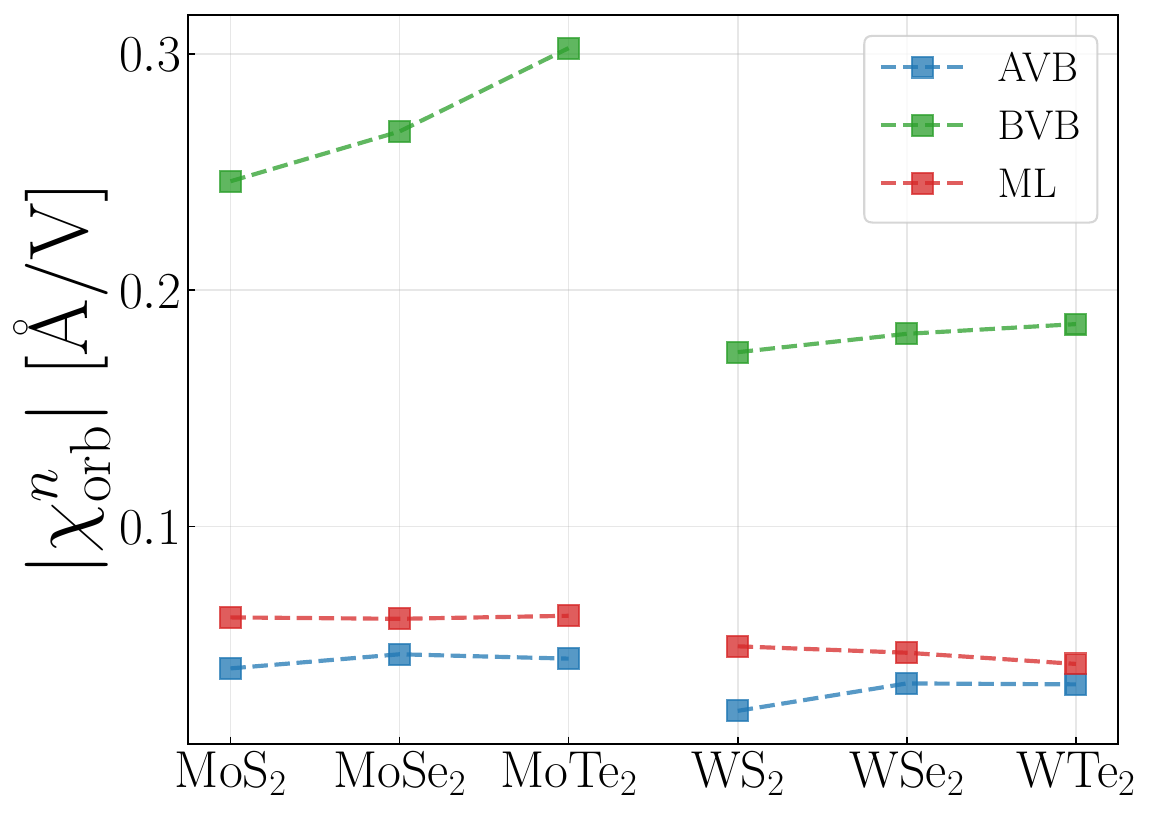}
    \caption{Orbital polarizability $\chi_{\mathrm{orb}}^n$ for the VB maximum of various TMD MLs, as well as the AVB and BVB states of the  TMD homo-BLs.}
    \label{fig:chi_orb}
\end{figure}

Since the dependence of the orbital polarization imbalance $\xi^n$ on the external field is nearly linear (see Supplementary Figure S1), we can define a meaningful derivative
\begin{equation}
   \chi_{\mathrm{orb}}^n = \frac{d\xi^n}{dE_{\mathrm{ext}}}.
\end{equation}
This quantity describes how the charge asymmetry, encoded in $\xi^n$, is affected by an external field. Conceptually, it is analogous to a polarizability, and we tentatively refer to $\chi_{\mathrm{orb}}^n$ as orbital polarizability. 
\Cref{fig:chi_orb} displays $\chi_{\mathrm{orb}}^n$ for various materials revealing that its magnitude is smallest in the AVB, intermediate at the VB maximum of the ML, and largest in the BVB, matching the hierarchy of $\lambda_\mathrm{R}^n$, shown in \cref{fig:rpar_E0} (a). 
Interestingly, W-based systems (right) show reduced $\chi_{\mathrm{orb}}^n$, 
explaining the behavior of $\lambda_\mathrm{R}^n$ when replacing Mo with W: Whereas heavier metal (W) has larger atomic SOC, the reduced polarizability limits the overall splitting. 

We thus propose  $\xi^n$  and   $\chi_{\mathrm{orb}}^n$ as efficient, generalizable descriptors for Rashba spin splitting, readily applicable to more complex stacks and heterostructures of two-dimensional materials. 

\subsection{Microscopic model} 

The behaviors of the orbital polarization imbalance and orbital polarizability underscore the central role of orbital asymmetry---particularly involving chalcogen $p_z$ orbitals---in determining both the magnitude and sign of Rashba splitting. To further deepen this understanding, we develop a microscopic model that elucidates the underlying mechanisms at the orbital level. Specifically, we incorporate SOC perturbatively within a tight-binding framework and compute the resulting spin splitting. This splitting incorporates naturally the asymmetry described by $\xi^n$, thus providing a direct qualitative explanation for the DFT trends.

We start from the description of the ML. Near the $\Gamma$-point, the top VB  of the ML is primarily composed of orbitals with $m_l=0$,  namely, the  $d_{z^2}$,  $p_{z_1}$ and $p_{z_2}$ orbitals from the metal, lower and upper chalcogens, respectively. Accordingly, we define the unperturbed state as:
\begin{equation}
\label{eq:psi0} 
|\psi^{(0)}\rangle\simeq\frac{1}{N}\left[\frac{1-\epsilon}{\sqrt{2}}\left|p_{z_{1}}\right\rangle-\frac{1+\epsilon}{\sqrt{2}}\left|p_{z_{2}}\right\rangle+b\left|d_{z^{2}}\right\rangle\right] \otimes|\sigma\rangle, 
\end{equation}
where $b$ is the relative $d_{z^2}$ contribution,  $\sigma$ denotes the spin degree of freedom and $N=\sqrt{1+b^2} + \mathrm O( \epsilon^2)$ is a normalization constant, while $|\epsilon|\ll 1$ captures the breaking of the $m_{z}$.    
Defined in real space, $\epsilon$ has similar field dependency as the orbital polarization  introduced above for Bloch states,  $\xi^{n}_{\mathrm{ML}}$. Namely, 
\[ |\xi^{n}_{\mathrm{ML}}| =c_{\mathrm{ML}}^n k\epsilon, \] 
where $c_{\mathrm{ML}}^n$ is an irrelevant constant for this discussion. We assume in the following: (i) the long-wavelength approach, $k \ll a^{-1}$, where $a$ is the lattice constant, (ii) the atomic SOC is small compared with other relevant energy scales (hopping, orbital energies, crystal field corrections, energy difference between orbitals or gap) and (iii) small orbital asymmetry due to an electric field, $|\epsilon|\ll 1$, as stated above.  
 
Rashba splitting arises at lowest order by hybridization of the $m_l=0$ states ($p_z$, $d_{z^2}$) with $m_l=\pm 1$ orbitals ($p_x$, $p_y$, $d_{xz}$, and $d_{yz}$), mediated by momentum-dependent hopping and atomic SOC. Without losing generality, we consider momentum along $k_x$ near $\Gamma$, $k_x\ll a^{-1}$, that is,  the mixing is limited to $p_x$ and $d_{xz}$ orbitals to lowest order. Two leading second-order processes contribute to spin splitting linearly in momentum and field:
\begin{align}
& \text{(i)}\quad \left|p_{z_i}\right\rangle \xrightarrow{\text{NN hopping}} \left|d_{xz}\right\rangle \xrightarrow{\text{SOC}} \left|d_{z^2}\right\rangle, \nonumber\\
& \text{(ii)}\quad \left|d_{z^2}\right\rangle \xrightarrow{\text{NN hopping}} \left|p_{x_i}\right\rangle \xrightarrow{\text{SOC}} \left|p_{z_i}\right\rangle.\nonumber
\end{align}
We focus first on the nearest neighbor (NN) hopping within the two-center approach,  introducing mixing with $m_l=\pm 1$ orbitals in reciprocal space:
\begin{align}
& \text{(i)}\quad |\psi_d^{(1)}\rangle = i \epsilon (k_x a) \frac{t_{z,xz}}{\varepsilon_{pd}} |d_{xz}\rangle\otimes|\sigma\rangle,\nonumber \\
& \text{(ii)}\quad |\psi_p^{(1)}\rangle = -i\epsilon (k_x a) b\frac{t_{x,z^2}}{\varepsilon_{pd}} (|p_{x_1}\rangle + |p_{x_2}\rangle)\otimes|\sigma\rangle,\nonumber
\end{align}
where $t_{z,xz}=3[\sqrt{3}n_z^2 V_{pd\sigma} +(1-2n_z^2)V_{pd\pi}]$ and $t_{x,z^2}=3[\sqrt{3}n_z^2 V_{pd\pi} +(1-2n_z^2)V_{pd\sigma}]$   are effective hopping amplitudes determined from Slater–Koster integrals $V_{pd\sigma}, V_{pd\pi}$ \cite{PhysRev.94.1498}, $n_z$ is the directive cosine along $z$  and $\varepsilon_{pd}$ the energy denominators of the intermediate virtual states. 

SOC then connects the electron spin and orbital degrees of freedom,   $\hat V_{\rm SOC} = \lambda_l\hat{\mathbf{L}} \cdot \hat{\mathbf{\sigma}} $, where $\hat{\mathbf{\sigma}}$ is the Pauli matrices vector, $\hat{\vec{L}} $ is the angular momentum operator given in the basis of direct atomic orbitals, and $\lambda_l$ is the angular momentum resolved atomic SOC strength, with $l = \{s,p,d,\dots \}$. This yields a spin-dependent correction to the energy, $\varepsilon^{(2)}_\alpha =\langle \psi^{(0)}_{\rm ML}|\hat V_{\rm SOC}|\psi_\alpha^{(1)}\rangle$:
\begin{align}
& \text{(i)}\quad \varepsilon^{(2)}_{d}= \frac{2\sqrt{3}b\epsilon}{N^2}k_xa\frac{t_{z,xz}}{\varepsilon_{pd}}  \lambda_dS_y \nonumber \\
& \text{(ii)}\quad \varepsilon^{(2)}_{p} =-\frac{2b\epsilon}{N^2} k_xa\frac{t_{x,z^2}}{\varepsilon_{pd}}  \lambda_p S_y,\nonumber
\end{align}
where $S_y$ is the spin component perpendicular to momentum, typical of a Rashba-type interaction. Adding both contributions, we obtain a linear-in-$k$ and linear-in-field splitting (recall that $\epsilon$ scales linearly with $E_{\mathrm{ext}}$), 
\begin{align}
& \Delta \mathrm{E}= \frac{2b k_xa}{N^2}\frac{(\sqrt{3}t_{z,xz}\lambda_d -t_{x,z^2}\lambda_p ) }{\varepsilon_{pd}} \epsilon.   \nonumber 
\end{align}

This can be generalized to a BL structure in the R$_{\mathrm{X}}^{\mathrm{M}}$/R$_{\mathrm{M}}^{\mathrm{X}}$ stacking. By inspection of the symmetry and orbital composition of the obtained DFT results, see \cref{fig:bands_intro.pdf}(d),(e), we have for the AVB at $E_\mathrm{ext} =0$:
\begin{align*}
 \psi^{(0)}_{\mathrm{A}}=&\frac{1}{N_\mathrm{A}}
\left[\frac{1-\epsilon_{\mathrm{L}_1}^\mathrm{A}}{\sqrt 2} \left|p_{z_{1}}\right\rangle-\frac{1+\epsilon_{\mathrm{L}_1}^\mathrm{A}}{\sqrt 2}\left|p_{z_{2}}\right\rangle+b_\mathrm{A}\left|d_{z^{2}_1}\right\rangle\right.  
\nonumber \\ 
&\left.-\frac{1+\epsilon_{\mathrm{L}_2}^\mathrm{A}}{\sqrt 2 }\left|p_{z_{3}}\right\rangle+\frac{1-\epsilon_{\mathrm{L}_2}^\mathrm{A}}{\sqrt 2}\left|p_{z_{4}}\right\rangle-b_\mathrm{A}\left|d_{z^{2}_2}\right\rangle\right], \nonumber 
\end{align*}
where $\left|p_{zi}\right\rangle$ with $i$=1,2,3,4 are the $p_z$ orbitals in ascending order along the stacking direction, $d_{z^2_{i}}$ are the metal contributions of the bottom ($i=1$) and top layers ($i=2$), respectively, and $\epsilon_{\mathrm{L}_i}^\mathrm{A} \propto |\xi_{\mathrm{L}_i}^\mathrm{A}|/k_x$ describe the break of symmetry of the AVB at the ML level. Note that the signs of $\epsilon_{\mathrm{L}_i}^{\mathrm{A,B}}$ are adapted to the DFT results of \cref{fig:bands_intro.pdf}(d),(e). Treating the MLs separately and repeating the calculations above, we obtain:

\begin{equation*}
\Delta \mathrm{E}^\text{AVB}=\frac{2b_{\mathrm{A}}k_xa}{N_\mathrm{A}^{2}}  \frac{(\sqrt{3}t_{z,xz}\lambda_{d} -t_{x,z^2}\lambda_{p} )}{\varepsilon_{p d}} ( \epsilon^\mathrm{{A}}_{\mathrm{L}_1}-\epsilon^\mathrm{A}_{\mathrm{L}_2}), 
\end{equation*}
where the individual Rashba contributions from each ML cancel partially,
which aligns with the concept of orbital polarization imbalance (recall that each ML has a charge asymmetry of different sign), that is, $\xi^{\mathrm{A}} \propto  \epsilon^\mathrm{{A}}_{\mathrm{L}_2}-\epsilon^\mathrm{A}_{\mathrm{L}_1}$. Moreover, the mirror symmetry is broken at $E_{\mathrm{ext}} =0$, that is, $\epsilon^\mathrm{A}_{\mathrm{L}_2}-\epsilon^\mathrm{A}_{\mathrm{L}_1}$ is in general finite at zero field, yielding a Rasbha spliting.  

For the BVB, we have:  
\begin{align*}
 \psi^{(0)}_{\mathrm{B}}=&\frac{1}{N_\mathrm{B}}
\left[\frac{1+\epsilon_{\mathrm{L}_1}^\mathrm{B}}{\sqrt 2} \left|p_{z_{1}}\right\rangle-\frac{1-\epsilon_{\mathrm{L}_1}^\mathrm{B}}{\sqrt 2}\left|p_{z_{2}}\right\rangle+b_\mathrm{B}\left|d_{z^{2}_1}\right\rangle\right.  
\nonumber \\ 
&\left.+\frac{1-\epsilon_{\mathrm{L}_2}^\mathrm{B}}{\sqrt 2 }\left|p_{z_{3}}\right\rangle-\frac{1+\epsilon_{\mathrm{L}_2}^\mathrm{B}}{\sqrt 2}\left|p_{z_{4}}\right\rangle+b_\mathrm{B}\left|d_{z^{2}_2}\right\rangle\right]. \nonumber 
\end{align*}
The situation is formally similar to the AVB case, with both ML contributions having opposite signs,  
\begin{equation*}
\Delta \mathrm{E}^\text{BVB}=-\frac{2b_\mathrm{B}k_xa}{N_\mathrm{B}^{2}}  \frac{(\sqrt{3}t_{z,xz}\lambda_{d} -t_{x,z^2}\lambda_{p} )}{\varepsilon_{p d}} ( \epsilon^\mathrm{B}_{\mathrm{L}_1}-\epsilon^\mathrm{B}_{\mathrm{L}_2}). 
\end{equation*}
However, the sign of the splitting is the opposite as for the AVB case (that is, the spin-ordering of the bands is reversed), which is consistent with the sign of the orbital polarization and with the numerical results at zero field (see \cref{fig:lambda_efield_bil_Se}).  
Moreover, the  difference of the orbital polarization of the MLs is larger for BVB (see \cref{fig:bands_intro.pdf}(e), schematically), resulting in a larger splitting for the BVB. 
 \begin{figure}[tbph]
    \centering
    \includegraphics[width=0.5\textwidth]{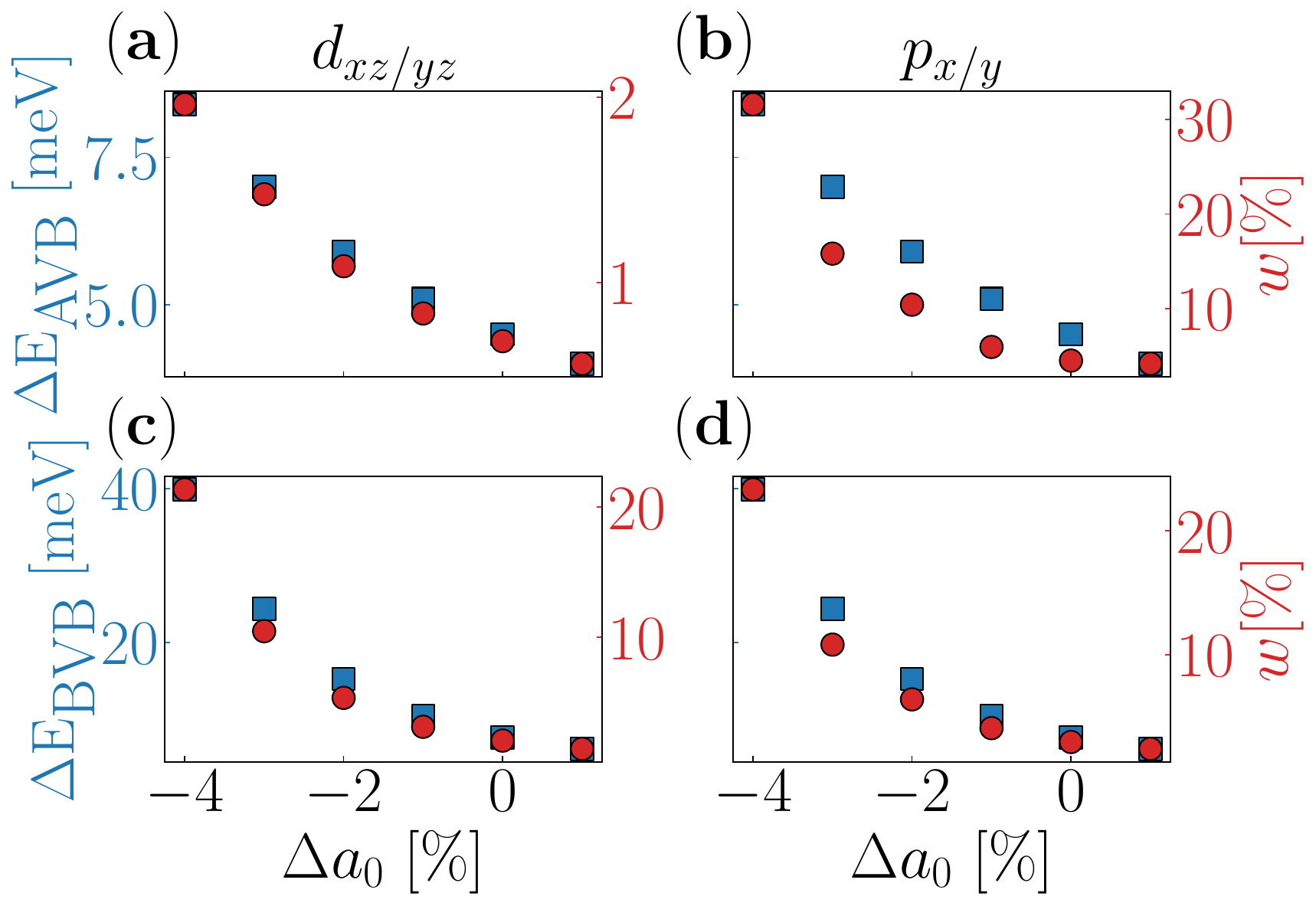}
    \caption{Rashba-splitting and orbital weight as a function of in-plane biaxial strain for BL MoSe$_2$ at $k={\mathrm{K}}/{16}$. The upper (lower) row shows the AVB (BVB) results, resolved for $d_{xz/yz}$ orbitals in the first column and $p_{x/y}$ orbitals in the second column.}
    \label{fig:compare_soc_weight_to_strain_MoSe2}
\end{figure}

To evaluate the role of orbitals involved in mechanisms (i) and (ii), \cref{fig:compare_soc_weight_to_strain_MoSe2} presents the Rashba spin splitting (shown in blue, left y-axis) alongside the percentage contributions of $d_{xz/yz}$ and $p_{x/y}$ orbitals (depicted in red, right y-axis) in the AVB (a),(b) and BVB (c),(d) as functions of in-plane biaxial strain.
These values are obtained from Wannierization of the DFT Bloch functions in the vicinity of the $\Gamma$-point, precisely at $k={\mathrm{K}}/{16}$.   
We observe that both the spin splitting and orbital weights of the in-plane orbitals ($d_{xz/yz}$ and $p_{x/y}$), show  a similar strain dependence. This similarity points to  the crucial role of these mediator orbitals, which, despite their low weight in the states of interest at zero strain, enable the transition required for Rashba SOC. It also underscores the importance of the mechanism (ii) involving the chalcogen SOC as a key contributor and explaining the pronounced increase in $\lambda_\mathrm{R}^n$ for heavier chalcogens. Finally, we note that the in-plane $d$-orbital weights are much higher in the BVB, explaining a larger spin splitting on this band. Recall that the BVB is energetically closer to other in-plane $d$-bands, enabling a more efficient hybridisation. 

\section{Conclusion}

In summary, we have carried out a systematic first-principles and microscopic analysis of Rashba spin splitting in monolayer and homo-bilayer MX$_2$ (M = Mo, W; X = S, Se, Te) TMDs, focusing on the top valence bands near the $\Gamma$-point. We obtained compact and consistent Rashba parameters across the TMD family within the linear-in-$k$ and linear-in-field regime. 

Although W-based compounds possess stronger atomic SOC than their Mo-based counterparts, this is  compensated by a reduced orbital polarization (i.e., a weaker layer-resolved charge asymmetry) in the relevant valence bands, resulting in Rashba coefficients of the same order of magnitude for Mo- and W-based systems. In contrast, heavier chalcogen atoms exhibit both enhanced atomic SOC and a larger orbital polarization, leading to a monotonic increase of the Rashba coefficient with chalcogen mass, independent of the transition metal.

While Rashba splitting in  R$_\mathrm{X}^\mathrm{M}$-stacked bilayers is allowed even at zero external field due to the intrinsic breaking of in-plane mirror symmetry, we find that the Rashba coefficient of the top valence band (AVB) is smaller than in the ML. This reduction is not obvious \textit{a priori}, since the BL hosts an internal polarization that already enables Rashba coupling in the absence of external bias. An even more striking effect emerges within the bilayer itself: the Rashba response differs strongly between the two top valence bands, with the AVB exhibiting a much weaker splitting than the next valence band (BVB) under identical symmetry conditions. To elucidate both trends, we introduce the band-resolved orbital polarization $\xi^n$ and its field derivative, the orbital polarizability $\chi^n_{\mathrm{orb}}$. We show that $\xi^n$ correlates directly with the Rashba coefficient and accurately predicts the critical electric field $E_0^n$ at which the spin splitting vanishes, demonstrating that Rashba physics in these systems is governed by internal orbital asymmetries rather than crystal symmetry alone.

A central outcome of this work is a microscopic model that explains Rashba spin splitting in terms of perturbative orbital mixing combined with atomic SOC. To our  knowledge, this is the first such model formulated for the $\Gamma$-point valence bands of TMDs. It identifies two dominant Rashba pathways, mediated by $p\leftrightarrow d $ orbital couplings, and naturally accounts for the distinct Rashba behavior of monolayers and bilayers.
By complementing the model with Wannier projections of the DFT Kohn–Sham states, we confirm that the leading first-order Rashba processes are governed by the transition-metal $d_{xz/yz}$ orbitals and the chalcogen $p_{x,y}$ ones. 
 
 The resulting zero-field Rashba parameters span approximately 1-200 meV$\mathrm{\AA}$ across the materials studied  (from WS$_2$ to MoTe$_2$, see Table 1 in the Supplementary Material). This range is comparable to, and in some cases exceeds, values reported for conventional Rashba systems such as Si-SiGe quantum wells ($\simeq$0.01 meV$\mathrm{\AA}$)  \cite{Prada2011}, oxide interfaces like  LaAlO$_3$/SrTiO$_3$  (10-50 meV$\mathrm{\AA}$)\cite{PhysRevB.87.161102},  semiconductor heterostructures including  InGaAs/InAlAs (40-70 meV$\mathrm{\AA}$) \cite{PhysRevB.39.1411,PhysRevLett.78.1335}   or AlGaN/GaN (81 meV$\mathrm{\AA}$ )\cite{Cho2005}.
In summary, by combining a quantitative first-principles analysis with a microscopic orbital model, this work clarifies how Rashba spin splitting in layered semiconductors is governed by internal orbital asymmetries. In particular, the orbital polarization imbalance and its polarizability serve as direct, physically intuitive measures of both the magnitude and tunability of the Rashba coupling. This insight enables predictive control via stacking configurations and external gating. These concepts naturally extend to more complex van der Waals systems, such as Janus layers and heterobilayers, where built-in asymmetry and interlayer hybridization can be engineered to tailor spin-orbit effects.

\section*{Acknowledgements}
The authors would like to thank Carl Emil Mørch Nielsen for fruitful discussions.
The project is supported by the Deutsche Forschungsgemeinschaft (DFG, German Research Foundation) within the Priority Program SPP2244 2DMP and by the Cluster of Excellence “Advanced Imaging of Matter” of the DFG -- EXC 2056 -- project ID 390715994. Calculations were carried out on Hummel funded by the DFG – 498394658.
\bibliographystyle{apsrev4-2}
\bibliography{references_main}

\end{document}